\documentclass[useAMS,usenatbib,a4paper]{mn2e}
\usepackage{graphicx,natbib,color}
			
\newcommand{\comment}[1]{}

\title[Open-loop tomography with ANNs]{Open-loop tomography with artificial neural networks on CANARY: on-sky results}

\author[J. Osborn et al.]{J. Osborn$^1$\thanks{E-mail: james.osborn@durham.ac.uk (JO)}, D. Guzman$^2$, F.~J. de Cos Juez$^3$, A.~G. Basden$^1$, T.~J. Morris$^1$,
\newauthor{E. Gendron$^4$, T. Butterley$^1$, R.~M. Myers$^1$, A. Guesalaga$^2$, F. Sanchez Lasheras$^3$,}
\newauthor{M. Gomez Victoria$^3$, M.~L. S\'{a}nchez Rodr\'{i}guez$^3$, D. Gratadour$^4$ and G. Rousset$^4$}\\	 
$^1$Department of Physics, Centre for Advanced Instrumentation, University of Durham, South Road, Durham, DH1 3LE, UK\\
$^2$Dept. of Electrical Engineering, Pontificia Universidad Catolica de Chile, Vicu\~{n}a Mackenna 4860, Santiago, Chile\\
$^3$Mining Exploitation and Prospecting Department, C/Independencia n¼13, University of Oviedo, 33004 Oviedo, Spain \\
$^4$LESIA, Observatoire de Paris, Section de Meudon 5, Place Jules Janssen, 92195 Meudon Cedex, France}
\begin{document}	

\maketitle		
\label{firstpage}

\begin{abstract}
We present recent results from the initial testing of an Artificial Neural Network (ANN) based tomographic reconstructor Complex Atmospheric Reconstructor based on Machine lEarNing (CARMEN) on Canary, an Adaptive Optics demonstrator operated on the 4.2~m William Herschel Telescope, La Palma. The reconstructor was compared with contemporaneous data using the Learn and Apply (L\&A) tomographic reconstructor. We find that the fully optimised L\&A tomographic reconstructor outperforms CARMEN by approximately 5\% in Strehl ratio or 15~nm rms in wavefront error. We also present results for Canary in Ground Layer Adaptive Optics mode to show that the reconstructors are tomographic. The results are comparable and this small deficit is attributed to limitations in the training data used to build the ANN. Laboratory bench tests show that the ANN can out perform L\&A under certain conditions, e.g. if the higher layer of a model two layer atmosphere was to change in altitude by $\sim$300~m (equivalent to a shift of approximately one tenth of a subaperture). 
\end{abstract}


\section{Introduction}

The next generation of large and extremely large telescopes require sophisticated Adaptive Optics (AO) instrumentation which exploit tomographic reconstruction algorithms in order to optimise the correction over the full field of view of the telescope\comment{ REF}. 

Open-loop tomographic AO systems such as Multi-Object Adaptive Optics (MOAO) \citep{Assemat07} instruments use several guide stars (natural and laser) distributed in the field to probe the turbulent atmosphere. The tomographic reconstructor uses this information to reconstruct the phase aberration along the line of sight to the scientific target, which is not necessarily along the same line as a guide star. MOAO systems include several of these target directions, each of which contain its own wavefront correcting device. MOAO is forced to operate in open-loop as each target direction requires its own reconstructed wavefront from the shared guide star wavefront sensors (WFSs). 

Most open-loop tomographic reconstructors require the contemporaneous atmospheric optical turbulence profile (i.e the strength of the optical turbulence as a function of the altitude) in order to optimise the correction. This is either measured independently by an external profiling instrument such as SLODAR \citep{Wilson02} or SCIDAR \citep{Vernin73}, or calculated directly from the WFSs \citep{Cortes12}. If the atmospheric optical turbulence profile was to change significantly during the astronomical observations the reconstructor would have to be updated in order to ensure that optimum performance was retained. Recent measurements of both the wind velocity and refractive index structure constant, $C_n^2(h)\mathrm{d}h$, altitude profile evolution throughout a night with a new SCIDAR instrument (Stereo-SCIDAR, \citealp{Shepherd13,Osborn13}) shows that both of these parameters can fluctuate significantly on the order of minutes.

The magnitude of the change in the optical turbulence profile that can be tolerated is not trivial to derive and depends on the specifications of the individual AO system.\comment{is a contentious issue.} This issue along with measurements of the temporal atmospheric variability will be presented in separate publications.

Learn and Apply (L\&A, \citealp{Vidal10}), is an open-loop tomographic reconstructor which actively learns the atmospheric profile. The measurements from all of the WFSs are combined and theoretical functions are used to recover the turbulence profile. This profile is then used to optimise the reconstructor. This method is extremely successful and has been implemented in the Canary MOAO demonstrator \citep{Gendron11}.

Canary is a flexible adaptive optics (AO) demonstration bench at the 4.2 m William Herschel Telescope (La Palma). Canary is modular by design and is ideally suited to testing and validating many novel ideas and concepts in the field of AO and in the wider field of astronomical instrumentation. In order to fully understand the instrument and the performance of all of the concepts and prototypes that will be tested on it, the bench contains an atmosphere and telescope simulator/calibration unit. Canary also contains a truth sensor, an additional on-axis WFS. This WFS is not used as an input to the tomographic reconstructor, but can be used to assess the performance of the AO system. It is located after the deformable mirror (DM) in the optical train and so can be used to measure the corrected wavefront, or by flattening the DM can be sued to measure the uncorrected on-axis wavefront.

In this paper we present the latest results from an on-going project to implement an Artificial Neural Network (ANN) as an open-loop AO tomographic reconstructor. ANNs are computational models inspired by biological neural networks which consist of a series of interconnected simple processing elements called neurons. Each neuron receives a series of data (input) from other neurons or an external source and transforms it locally using an activation or transfer function. These output data are then transferred to other neurons with different weights and the cycle continues until the output neurons are reached. The network needs to be trained before it can be used. During the training, the weights are changed to adopt the structure of a determined function, based on a series of input-output data sets provided. Although each individual neuron implements its function slowly and imperfectly, the whole structure is capable of learning complex functions and solutions quite efficiently \citep{Reilly90}.

\comment{ANN review?}
ANNs have been applied to the field of AO. Previously, this has been concentrated on wavefront sensing algorithms. \cite{Montera96} applied an ANN to centroid images in a Shack-Hartmann wavefront sensor to estimate the local slopes. They found that although the ANN was no better than the standard `Centre of Gravity' type approaches, however, the ANN was better at estimating the magnitude of the wavefront sensing error. In addition, \cite{Angel90}, \cite{Sandler91} and \cite{Lloyd-Hart92} successfully implemented an ANN for wavefront sensing in the focal plane.

\comment{\comment{form optics express paper}
ANNs have been applied to the field of AO in the past. Angel \textit{et al.} (1990) \cite{Angel90}, Sandler {\it et al.} (1991) \cite{Sandler91} and Lloyd-Hart {\it et al.} (1992) \cite{Lloyd-Hart92} present successful results using neural networks for wavefront sensing in the focal plane. Montera {\it et al.} (1996) \cite{Montera96} experimented with an ANN to reduce WFS centroiding error and to estimate the Fried parameter $r_{0}$ and the WFS slope measurement error. They found that the ANN performed as well as but not better than a standard linear approach for estimating the WFS slopes and for the estimation of the Fried parameter, r0, however the ANN was very good at estimating the WFS slope measurement error. ANNs have also been investigated for spatial and temporal predictions of the slope measurements. Lloyd-Hart \& McGuire (1995) \cite{Lloyd-Hart95} use an ANN to make a temporal prediction of the WFS slopes. The AO latency \comment{(bandwidthXX) }is then reduced allowing for a better correction. Weddell \& Webb (2006, 2007) \cite{Weddell06, Weddell07} developed this idea and used off-axis WFS measurements to temporally predict the on-axis slopes. However, this was limited to low-order Zernike modes (tip/tilt) only. More recently neural networks have  been used to model open loop DMs for MOAO \cite{Guzman10}. An accurate DM model is required for open-loop AO as the DM is not seen by the WFS.
}

The purpose of this project is to develop an open-loop tomographic reconstructor which is entirely insensitive to changes in the atmosphere optical turbulence profile. In \cite{Osborn12} we demonstrated an ANN implementation of an open-loop tomographic reconstructor, called `CARMEN', in a Monte-Carlo simulation. This simulation also had an implementation of the Learn and Apply (L\&A) and a simple least-squares matrix-vector-multiplication reconstructor. We demonstrated that CARMEN had the potential to attain a better performance than the other two reconstructors. This was true in the case when it was compared to fully optimised reconstructors and in the case when the atmosphere dynamically changed during the simulation duration. However, it should be noted that the ANN reconstructor was trained with the same simulation, albeit with different data. This is not possible for the on-sky implementation. The ultimate goal is to develop a reconstructor which is insensitive to changes in the atmospheric optical turbulence profile. Therefore, we do not wish to train the ANN with on-sky data as this would result in the reconstructor learning the concurrent profile. Instead we propose to train the ANN off-line on an AO calibration bench.

Canary is an ideal AO demonstrator on which to develop this reconstructor. The Canary calibration unit is used to generate the training data sets for CARMEN. We need to generate the training datasets on the same bench as we intend to use on-sky. This is because the neural networks, as with any reconstructor, will be sensitive to the relative alignment errors of the wavefront sensors.

In section~\ref{sect:ANN} we briefly describe the ANN. Section~\ref{sect:training} describes the training method, section~\ref{sect:implem} the ANN implementation into the Canary control system, and in section~\ref{sect:results} we show the results of the ANN reconstructor with both bench data and on-sky. We discuss the results in section~\ref{sect:discussion} and conclude in section~\ref{sect:conclusion}.

\section{Artificial Neural Networks}
\label{sect:ANN}

A characteristic of the ANN is its inherent ability to generalise. Once trained, the network is able to produce an optimised output based on previously unseen data \citep{Rooji96}. Moreover, it has been shown that ANNs perform well on data that are noisy, imprecise, and with incomplete observations\citep{Kasabov96, Osborn12}. At this juncture it should be noted that while the mathematical content of ANNs may be complex, the underlying model is basic in comparison to the massive computational power of the biological neuron \citep{Gurney97}. Nonetheless, when compared with traditional statistical predictive techniques ANNs have shown promising results \citep{Hua96}, hence their application in this instance.

The specific ANN model adopted for this work was the multilayer perceptron (MLP) with feedforward architecture, using the backpropagation training algorithm during a supervised training process. Model development was performed in R\comment{ (www.r-project.org)} using the {\textsc amore}\comment{ (http://cran.r-project.org/web/packages/AMORE/)} package. 

An MLP maps sets of input data onto a set of appropriate outputs \citep{Gurney97}. The MLP includes an activation function in each of the neurons. This activation function defines whether or not a particular neuron activates, or fires, given an input signal. \comment{A linear activation function is a simple on-off mechanism to determine whether or not a neuron fires. }For an MLP containing a linear activation function, it can be shown with linear algebra that any number of layers can be reduced to a standard matrix-vector multiplication input-output model. MLPs can also contain non-linear activation functions on each neuron. These functions are developed to model the frequency of action potentials, or firing, of biological neurons in the brain. This function is modelled in several ways, but must always be normalisable and differentiable.

The most popular activation function used in current applications is the hyperbolic tangent sigmoid function, this can be described as \citep{Gurney97,Haykin08},
\begin{equation}
\mathcal{F}_\mathrm{i} = \frac{2}{1+\exp{(-2v_i)}}-1,
\label{eq:sig}
\end{equation}
in which the function is a hyperbolic tangent which ranges from -1 to 1. $v_i$ is the weighted sum of the input synapses of the $i$th node (neuron). More specialised activation functions include radial basis functions which are used in another class of supervised neural network models, but were found not to be required here.

\begin{figure}
	\centering
	\includegraphics[width=\columnwidth]{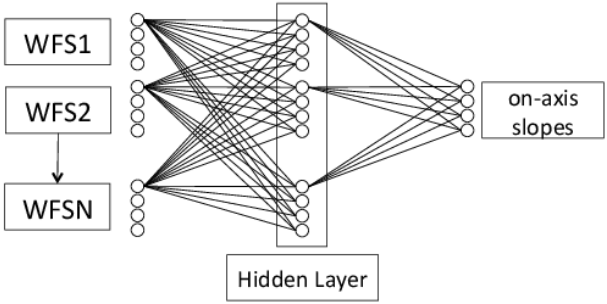}
	\caption{A simplified network diagram for CARMEN. All of the slopes from the WFS are input to the network. They are all connected to every neuron in the hidden layer by a synapse. Each neuron in the hidden layer is then connected to every output node. CARMEN will output the predicted on-axis wavefront slopes for the target direction. Each of the synapses has a weighting function. At run time the inputs are injected into the network which is then processed by the different weighting functions generating a response. In the diagram only a few of the synapses are shown for clarity.}
	\label{fig:net}
\end{figure}

The MLP consists of one input and one output layer with one or more intermediate (or hidden) layers of nodes with a nonlinear activation function (figure~\ref{fig:net}). Each node in one layer connects with a certain weight $w_{ij}$ to every node in the following layer. Learning occurs in the perceptron by changing connection weights (or synaptic weights) after each piece of data is processed based on the amount of error in the output compared to the expected result. This is an example of supervised learning and it is performed through back propagation, a generalisation of the least mean squares algorithm in the linear perceptron. We represent the error in output node $j$ in the $n$th data point by $e_j(n) = \left(d_j(n) - y_j(n)\right)^2$, where $d$ is the target value and $y$ is the value produced by the perceptron. We then make corrections to the weights of the nodes based on those nodal errors which minimise the energy of error in the entire output, given by,
\begin{equation}
\epsilon(n)=\frac{1}{2}\sum{\left(d_j(n) - y_j(n)\right)^2}
\end{equation}

A key summary of the model topology, or architecture, of CARMEN is provided in Table 1. The input layer is constructed of 504 input nodes (3 natural guide stars + 4 laser guide stars, all with $7\times7$ subapertures, resulting in 36 unvignetted subapertures per wavefront sensor, in x\&y). The output layer consists of 72 nodes which describes the open loop slopes for the on-axis target. The hidden layer consists of 504 nodes to match the input layer. Using more than one hidden layer had no discernible benefit to the model prediction accuracy. The type of transfer function used between the input to hidden and hidden to output layers was Òhyperbolic tangentÓ. The topology of the model presented here was determined largely by trial and error. Of the various models developed, this study reports on findings from the most accurate predictor model. For more details of the neural network we direct the reader to our previous work in \cite{Osborn12}
\begin{table*}
\centering
\caption{CARMEN model topology}
\begin{tabular}{ll}
\hline
Parameter & Value\\
\hline
Type of input & Continuous\\
Type of output & Continuous\\
Transfer function & tanh \\
Network connectivity & Fully connected\\
Learning Algorithm & Momentum\\
Learning rate coefficient & Input to hidden layer: 0.10\\
 & Hidden to output layer: 0.05\\
Number of hidden layers & 1\\
\hline
\end{tabular}
\end{table*}

\section{Training}
\label{sect:training}
ANNs are trained by exposing them to a large number of inputs together with the desired output. In theory this training data should cover the full range of possible scenarios. When the ANN is confronted with a superposition of a number of the independent training sets it can then predict an output by combining a number of the synaptic pathways. In this way we do not need to train the ANN with every possible turbulent profile but just a basis set from which it can assemble its own approximation.

We have used the Canary calibration bench to generate a training dataset. This bench consists of deployable four natural and three laser guide stars and two phase screens. The turbulence strength is distributed between these two phase screens with a ratio of 0.7:0.3. Initially we attempted to only use the stronger phase screen for the training (with a measured $r_0 = 0.25$~m). However, it became apparent that this phase screen alone did not have enough phase variability (i.e. was not big enough) for the statistics to converge for a suitable training set. If we do not train with sufficient variability in the input phase then the performance of the reconstructor is severely compromised \citep{Osborn12}. Therefore, for the training, we use the two available phase screens and place them as close together as possible. We then counter rotate them at different angular velocities to increase the variability, or independent realisations, of the phase that we measure. We place the phase screens at the ground and take 10000 iterations of WFS slopes, the angular velocities are defined so that the system is exposed to all possible combinations of the two phase screens. We then move the two layers up through the simulated atmosphere space together in small increments. The dataset then includes the influence of turbulent layers at all possible altitudes.

We train CARMEN to reconstruct the on-axis target slopes (i.e. the slopes that an on-axis WFS would measure if one were available) regardless of atmospheric turbulence profile. The input to the reconstructor will be the measured off-axis slopes from the guide star WFSs. The output of CARMEN will be the open loop slopes predicted for the on-axis target, which can be converted into DM commands with the truth sensor control matrix. This will result in a tomographic reconstructor that is stable even in dynamic atmospheric conditions. 

The training must be performed whilst the bench is set to have the same asterism that will be used on-sky. For this reason, a different reconstructor is required for each potential asterism. It is possible that the reconstructor for the laser guide stars could be separated from that of the natural guide stars, effectively resulting in two reconstructors. As the asterism of the laser guide stars for each system is fixed, a single reconstructor can always be used for the laser guide star reconstructor. For the tests we used Canary asterism 34, the guide star parameters are defined in table 2. The laser guide stars are positioned at the corners of a square, centred on the on-axis target star, with sides of length 37 arcseconds.
\begin{table*}
\centering
\caption{Canary asterism 34 parameters}
\begin{tabular}{llll}
\hline
Star & X (arcseconds) & Y (arcseconds) & Magnitude (in V band)\\
\hline
on-axis & 0 & 0 & 9.7 \\
natural guide star 1 & -7.2 & -20.2 & 9.2\\
natural guide star 2 & -36.0 & 53.3 & 9.7\\
natural guide star 3 & 54.0 & 4.3 & 11.6\\
laser guide star 1 & 18.5 & 18.5 & -\\
laser guide star 2 & 18.5 & -18.5 & -\\
laser guide star 3 & -18.5 & 18.5 & -\\
laser guide star 4 & -18.5 & -18.5 & -\\
\hline
\end{tabular}
\end{table*}

\section{Implementation}
\label{sect:implem}

Canary uses the Durham AO real-time controller (DARC) \citep{Basden11} to provide real-time actuator control
in response to WFS inputs.  This control system is modular, allowing
different algorithms, WFSs and deformable mirrors to be integrated with the real-time control system.
We have developed an ANN reconstruction module (written in the C
language) for DARC which takes advantage of the pipelined
architecture, minimising latency between the last WFS pixel
received and commands being sent to the DMs.  DARC modules are
dynamically loadable, enabling fast switching of control algorithms,
including when the AO loop is engaged.  We are therefore able to
compare traditional matrix-vector wavefront reconstruction approaches
with our ANN implementation with very little delay.

\subsection{DARC module design}
An ANN can be represented by a sequence of matrix-vector
multiplications interspersed with addition of a bias term, also derived from the training process and used to apply an offset to the neuron values at each layer, and a non-linear mapping:
\begin{equation}
x_\mathrm{i+1} = \mathcal{F}_\mathrm{i}\left( \textbf{M}_\mathrm{i} \cdot \textit{\textbf{x}}_\mathrm{i} + \textit{\textbf{b}}_\mathrm{i}  \right),
\end{equation}
were, $x_\mathrm{i}$ is the state vector after the
$\mathrm{i}^\mathrm{th}$ stage of the ANN, $M_\mathrm{i}$ is the
matrix corresponding to this stage, $b_\mathrm{i}$ is a vector bias
term and $\mathcal{F}_\mathrm{i}$ is the activation function for this stage.

We use a three-stage ANN for Canary, although our real-time
implementation will allow for an arbitrary number of stages.  The
first stage maps WFS slopes (504 from the seven Canary WFSs) to an
intermediate layer, which is then mapped to a layer representing
slopes as would be seen by an on-axis WFS with 72 slope measurements.
Finally, a linear stage is used to map this to DM commands using a
closed-loop control matrix (which is not part of the ANN learning).
We have the option to use intermediate stages with a Sigmoid
activation function, (equation~\ref{eq:sig}), or with a linear
activation function, allowing investigation of the ANN performance.\comment{  Our
DARC module also has the ability to use different activation
functions for different parts of each intermediate vector, however we
did not use this facility and do not report on it here.
When performing the intermediate matrix-vector multiplications, we
currently use dense matrices.  However, should there be a need
(e.g.\ for higher order systems), it would be trivial to implement
sparse matrix algebra if appropriate.}

\subsection{Utilisation of pixel streams}
A key feature of DARC is the ability to work with pixel streams
rather than image frames. The processing of pixels is performed as
they are delivered to the real-time control system, rather than having to wait for a
whole frame to arrive.  This is instrumental in delivering low
latency, and thus, improved AO performance.  Using the standard DARC
matrix-vector reconstruction module, partial DM commands can be
computed once enough pixels for a given sub-aperture have been
delivered (this sub-aperture is calibrated, slopes computed, and
partial reconstruction carried out).  For the ANN module, this is not
possible, since to progress from the first layer of the ANN to the
following layers, all slope measurements must be known.  However, our
implementation allows us to begin to process pixels as soon as they arrive at the real-time 
control system, rather than waiting for a whole image frame to arrive. As soon as enough pixels for a given sub-aperture have been captured, they 
are calibrated, and the slope measurements for this sub-aperture computed.  
Then, these slope measurements are used to perform a partial 
multiplication with the first ANN stage matrix, and as more slope 
measurements become available, the output of the first ANN stage is built 
up.  Finally, once all pixels have arrived, the output of the first ANN 
stage is then complete, and passed on to further ANN stages. This allows the first ANN stage to make use of the pixel
stream, and since this involves the largest matrix (a factor of seven
larger than the second stage matrix in the Canary case), we therefore
retain most of the benefit of the DARC pixel stream architecture. As a demonstration, the computation time for the L\&A and the ANN reconstructor was found to be $0.68\pm0.02$~ms and $1.01\pm0.01$~ms respectively. \comment{Therefore, the computation time is longer for the ANN, as expected, however it is still below the expected atmospheric coherence time of 3 to 5~ms.}

Being based on matrix-vector multiplications, the ANN module is a key
candidate for implementation on graphical processing unit hardware allowing operation for extremely large telescope (ELT) scale systems.

\section{Results}
\label{sect:results}
\subsection{Bench validation}
To validate the ANN tomographic reconstructor, CARMEN, we place the stronger of the two phase screens at the ground, $PS_1$. This phase screen is fixed to an altitude of zero for all tests as we assume that the surface turbulent layer is always present \citep{Osborn10}. The second phase screen, $PS_2$, is positioned at altitude to represent a high turbulent layer. We then move $PS_2$ to several different positions, corresponding to altitudes in the range $H_0-2000$ to $H_0+2000$, where $H_0$ is the altitude at which the L\&A tomographic reconstructor was optimised. The combined Fried parameter, $r_0$, a measure of the integrated atmospheric optical turbulence strength, for the two phase screens was $\sim$0.16~m.

\begin{figure}
	\centering
	\includegraphics[width=\columnwidth]{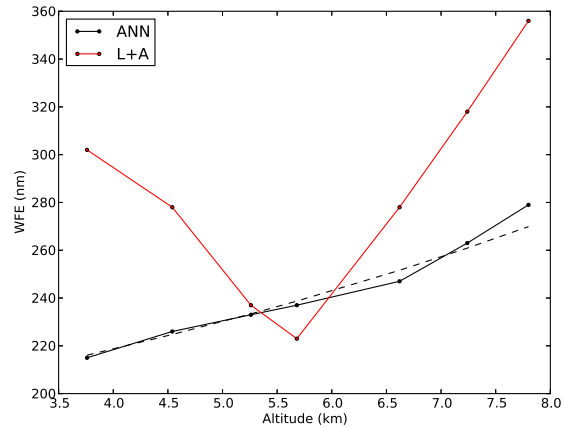}
	\caption{Residual WFE for L\&A and ANN tomographic reconstructors on the CANARY calibration bench with the high altitude turbulent phase screen ($PS_{2}$) position at the given altitude. The dashed line shows the expected performance of the ANN as a function of overlap of the projected pupils.}
	\label{fig:ANN_bench}
\end{figure}

Figure~\ref{fig:ANN_bench} shows the results from this experiment. We see that the performance is optimised for L\&A at an altitude of 5.7~km. As $PS_2$ is moved away from this altitude the performance degrades. The curve is not symmetric due to the first phase screen, $PS_1$, at the ground and due to the reduction in overlap of the projected pupil inducing further errors at increasing altitudes. The performance of CARMEN is approximately linear with increasing altitude. The reduction in performance is, theoretically, based on the reduced overlap of the projected pupils at altitude. The fraction of overlapping area of two full disks \comment{(not including secondary obscuration, this needs including) }separated by a distance x in diameter units where, $x=r/D$, is $f(x) = \arccos{(x)} - x(1-x^2)^{1/2}$. The residual error is proportional to $1/f(x)$. We see that that performance of CARMEN does follow this trend, which demonstrates that the reconstructor has been generalised to perform regardless of input atmospheric turbulence profile.

By comparing the performance of the L\&A and CARMEN reconstructors we see that a fully optimised L\&A reconstructor is indeed better than the generalised ANN reconstructor. However, if the high altitude turbulent layer was to change in altitude by $\sim$300~m and the L\&A algorithm is not re-optimised then the two reconstructors are equal. Beyond this altitude range CARMEN outperforms the L\&A reconstructor. For an asterism of $50^{\prime\prime}$ and 7 subapertures across a 4.2~m pupil (the simulated parameters of the optical bench) this altitude change corresponds to a shift of one tenth of a subaperture.

\subsection{On-sky validation}
During the nights of the 22nd and 24th July 2013 the bench trained ANN tomographic reconstructor was implemented on-sky. The first of these nights was spent calibrating the ANN reconstructor, involving developing the optimum routine for calculating the static aberrations and the optimum gain. 

Canary was operated for short bursts of $\sim$30~s with active switching between L\&A and CARMEN tomographic reconstructors. This methodology was used to prevent bias in the results by using different reconstructors at different times during changeable conditions. Due to time constraints only 36 exposures were made with each reconstructor and the Strehl ratio was recorded from the Canary science camera \comment{<REF>}in the H band. 

\begin{figure}
	\centering
	\includegraphics[width=\columnwidth]{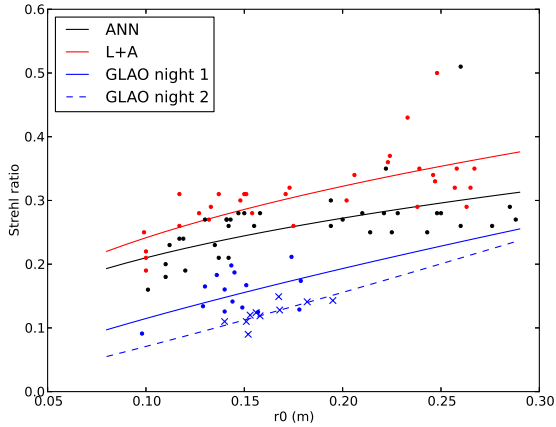}
	\caption{On-sky Strehl ratio (in H-band) achieved with the ANN reconstructor and with L\&A as a function of $r_0$. The reconstructors were interlaced temporally to prevent biasing due to changing conditions. The solid line indicates the least squares fit to the data with a power law model ($y=\alpha x^{\beta}$). We see that the results from L\&A is approximately 5\% better than that of CARMEN. Also shown is the performance of GLAO from the same nights, showing that the correction of L\&A and CARMEN was indeed tomographic.}
	\label{fig:strehl}
\end{figure}

Figure~\ref{fig:strehl} shows the Strehl ratio obtained with the two reconstructors as a function of $r_0$. $r_0$ is estimated by fitting the theoretical variances of a Zernike decomposition of the Kolmogorov power spectrum to those of the reconstructed wavefront from the WFS slopes \citep{Gendron11}. The L\&A reconstructor achieves a mean Strehl ratio of 0.31 $\pm$ 0.06, where the error given is the standard deviation. CARMEN achieved a mean Strehl ratio of 0.26 $\pm$ 0.06. This shows that CARMEN is capable of attaining a similar performance as L\&A on-sky. From the standard deviation of the data and by examining the figure we see that there is significant overlap in the results. We also present the results from Ground Layer Adaptive Optics (GLAO) measurements on the same nights (separated into night 1, 2013/07/22, and night 2, 2013/07/24, as the performance of GLAO was different for each night). GLAO is type of AO correction which only corrects for the ground layer of turbulence. We see that the performance of tomographic reconstructors (L\&A and CARMEN) is better than that of GLAO showing that the correction is indeed tomographic.

\begin{figure}
	\centering
	\includegraphics[width=\columnwidth]{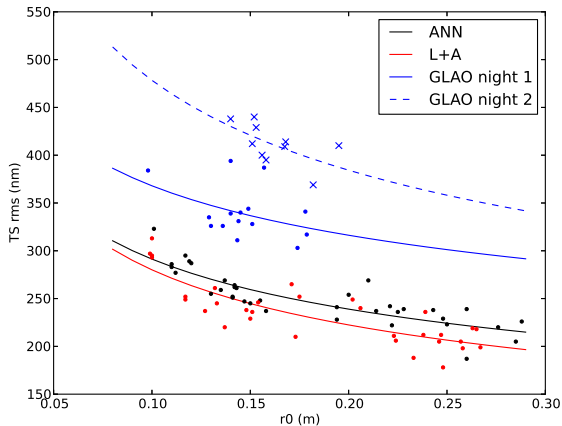}
	\caption{On-sky TSrms achieved with the ANN reconstructor and with L\&A as a function of $r_0$. The reconstructs were interlaced temporally to prevent biasing due to changing conditions. The solid line indicates the least squares fit to the data. We see that the mean difference in residual wavefront error from L\&A is approximately 15~nm rms lower than that of CARMEN. Also shown is the performance of GLAO from the same nights, showing that the correction of L\&A and CARMEN was indeed tomographic.}
	\label{fig:TSrms}
\end{figure}

\comment{
It is worth noting that during testing on the first night CARMEN a achieved the highest Strehl ever recorded on Canary, and hence any tomographic (IS THIS CORRECT?) instrument, of 51\%. Unfortunately, we are unable to compare this to the L\&A reconstructor as no data in this mode was recorded at that time.
}
In addition to the Strehl ratio we can also analyse the results in terms of residual wavefront error using WFS data from the Canary truth sensor (TS rms). Figure~\ref{fig:TSrms} shows the residual wavefront error again as a function of integrated $r_0$. A least squares fit to the data with a power law model ($y=\alpha x^{\beta}$) indicates a mean deficit of approximately 15~nm in the performance of CARMEN in comparison to L\&A for measured $r_0$ values between 0.08~m and 0.29~m. 

\begin{figure}
	\centering
	\includegraphics[width=\columnwidth]{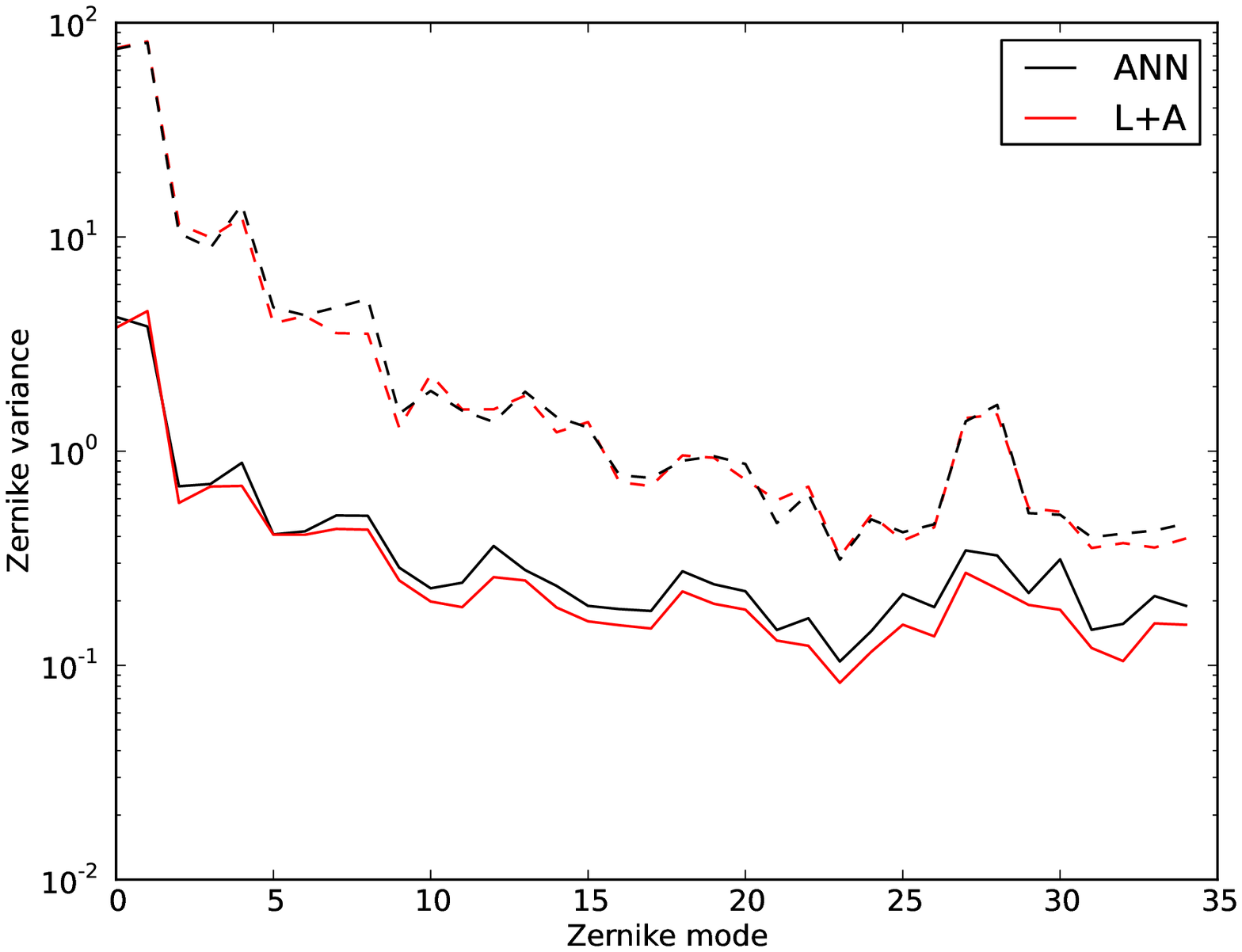}
	\caption{Zernike variance for the ANN and the L\&A reconstructor. The dashed lines indicate the Zernike variances for the uncorrected wavefront (calculated as the mean variances for the three off-axis natural guide stars) and the solid line indicates the corrected variances from the truth sensor.}
	\label{fig:Z_comp}
\end{figure}

\comment{

<T.~M. write something on quasi-static aberrations>
\comment{Quasi-static aberration which change over timescales longer than that of a single dataset are difficult to correct. Using this truth sensor it is possible to calculate the quasi-static residual aberration after the AO correction. As this aberration is static over the course of a dataset but changes over the duration of several datasets it needs to be included into the reference positions for the WFSs.}

If we subtract this quasi-static aberration term in an attempt to isolate the tomographic error <Do we not need to subtract actual aberrations and not just rms errors?> then we see that the performance of the ANN and of L\%A become very similar (figure~\ref{fig:TSrms_abstat}). This is because the static aberrations for the ANN are higher than that of L\%A. The static aberrations were measured in L\&A mode, effectively biasing the performance towards this mode of tomography.
\begin{figure}
	\centering
	\includegraphics[width=\columnwidth]{TSrms_abstat}
	\caption{}
	\label{fig:TSrms_abstat}
\end{figure}

}
By analysing the Zernike modal decomposition for each reconstructor we can attempt to understand the source of the discrepancy in performance. Figure~\ref{fig:Z_comp} shows the uncorrected Zernike variances (mean of the off-axis natural guide stars Zernike variances) and the corrected Zernike variances from the truth sensor. We find that the difference between the two reconstructors is greatest at higher order Zernike modes.

\comment{
\begin{figure}
	\centering
	\includegraphics[width=\columnwidth]{Z_ratio}
	\caption{}
	\label{fig:Z_ratio}
\end{figure}
}
\section{Discussion}
\label{sect:discussion}
\comment{At the time that these experiments were being performed we have concurrent information of the atmospheric profile of the optical turbulence from a Stereo-SCIDAR instrument mounted on the JKT \citep{Osborn13}. This data shows that we had three strong layers in the atmosphere at the time, positioned at 0~km, $\sim$1~km and  $\sim$3.5~km. Both of these layers are low altitude, explaining why the MOAO results is very similar to that of SCAO.}

\comment{<We need to put 15nm into context>}

The on-sky performance of CARMEN does not match that of the optimised L\&A. Although the results are similar, this small disparity mirrors the difference observed with the bench tests and could be caused by the finite number of independent phase realisation of the phase screens used for the training process. If we could further increase the bench phase variability we can expect the performance of CARMEN to improve. This is because, if there is not enough variability in the training data, then the measured power spectrum of the phase will not have converged and will be erroneous. This error is then imprinted into the ANN, which will attempt to force the output to match this erroneous power spectrum.

Bench tests showed that placing the two phase screens together and counter rotating provided better results than simply using one phase screen alone. The reason for this is that by using both phase screens we can increase the variability and the strength of the turbulence for the bench training. Using two phase screens induces further errors as there is a finite separation between them. It follows that the phase screens will therefore be conjugate to different altitudes on the calibration bench. We estimate that the two phase screens are separated by approximately 500~m (5~mm on the bench). The trained dataset will therefore contain information on double layers of separation 500~m. This will inevitably lead to errors in the reconstruction of single layers.

\comment{This means that when we train the network we are effectively training the ANN with two different sets of asterisms. For an asterism with a guide star which is 50$^{\prime\prime}$ off-axis the higher turbulent layer will appear to be from a light-cone offset by 10~cm from this. For a turbulent layer at 10~km it will appear as though the guide star was actually at 52.5$^{\prime\prime}$. This is shown in figure~\ref{fig:training_alt_offset}.
\begin{figure}
	\centering
	\includegraphics[width=\columnwidth]{training_alt_offset.eps}
	\caption{The separation of altitude, $\delta h$, of the two phase screens, $PS_1$ and $PS_2$, during training induces an apparent offset in the position of the slopes for one of them.}
	\label{fig:training_alt_offset}
\end{figure}
}

Errors in the conjugation altitude and lateral positions of the guide stars will mean that the geometry of the light cones on the bench will be different to that on-sky, inducing errors in the reconstructed on-axis slopes. This will have an effect on the beam overlap as a function of altitude and effectively mean that some WFS will see the turbulence further away than others. 
\begin{figure*}
	\centering
	\includegraphics[width=2\columnwidth]{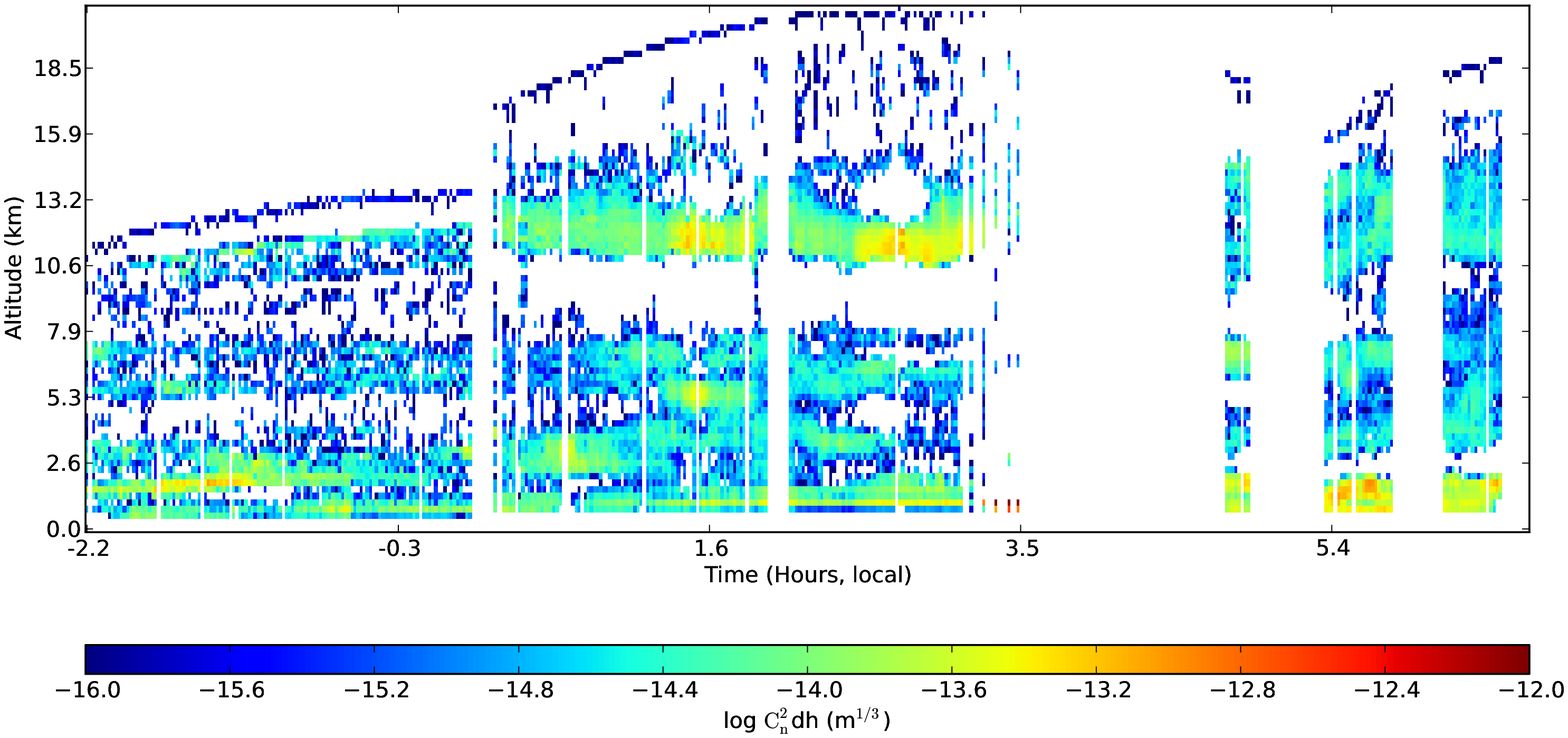}\\
	\vspace{-23mm}
	\includegraphics[width=2\columnwidth]{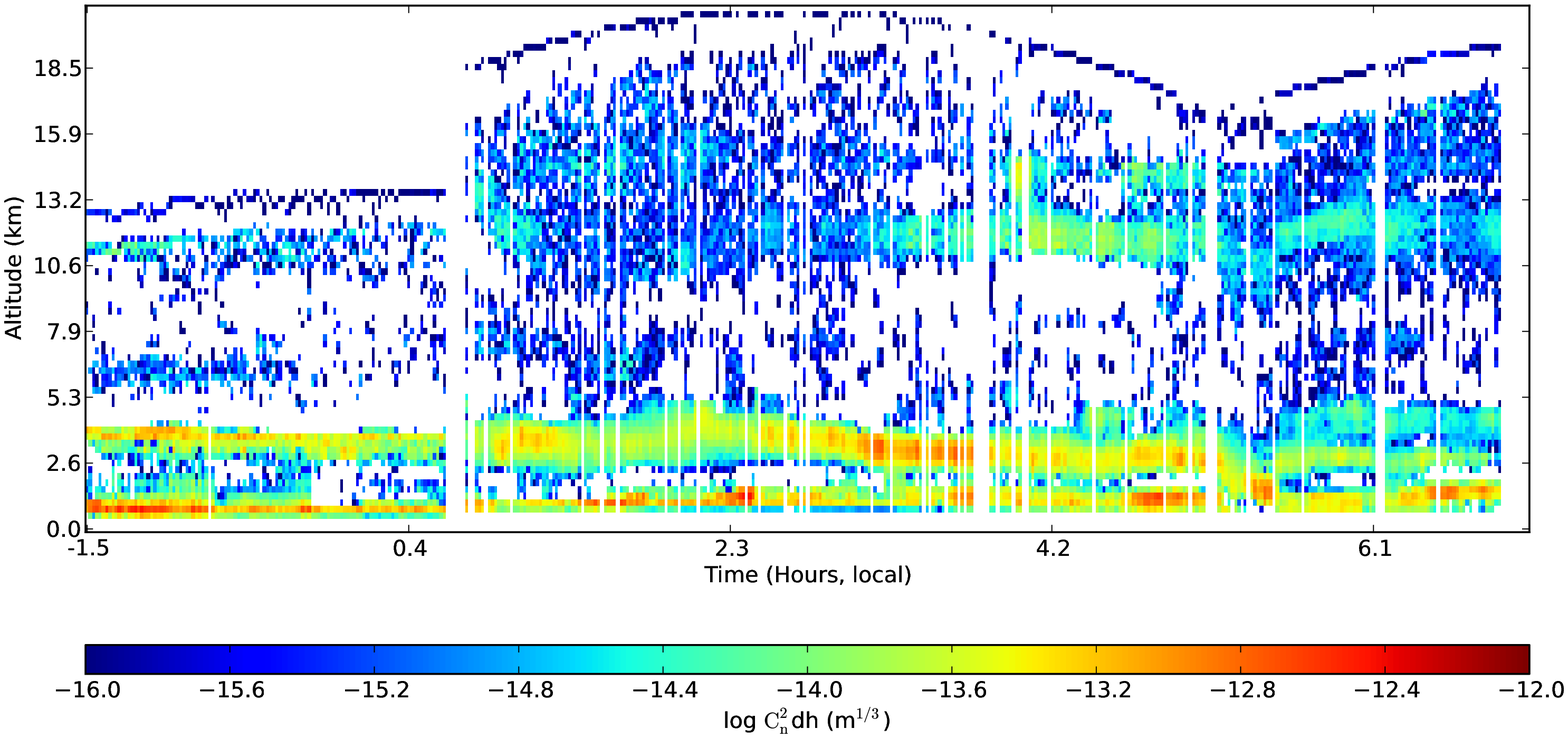}
	\caption{Atmospheric optical turbulence profiles for the first night (2013/07/22) and the second night (20130724) of CARMEN tests form Stereo-SCIDAR. The z-scale indicates the strength of the optical turbulence at a given time and altitude. On both nights the reconstructor tests were implemented at approximately 0030 to 0200. The GLAO data was taken at approximately 2230 -- 2240 and 0530 -- 0550 on the first night and 2240 -- 2310 on the second night.}
	\label{fig:SS_profs}
\end{figure*}

In addition, turbulence above the maximum altitude used in the training will not be corrected and will therefore reduce the performance of CARMEN. Due to limitations with the current bench used to generate the training data, the maximum altitude to which we can place a phase screen is approximately 6.5~km. Therefore, in its current format CARMEN is unable to correct for turbulence above this altitude. Concurrent turbulence profiles from an external Stereo-SCIDAR  instrument \citep{Shepherd13,Osborn13} on the Jacobus Kapteyn Telescope, La Palma, (figure~\ref{fig:SS_profs}) demonstrates that during testing there was approximately 15~\% of the integrated turbulence strength above this altitude on the first night and approximately 5~\% on the second. This high altitude turbulence would certainly reduce the performance of CARMEN. For this reason it is important to improve the training to include altitudes up to the maximum altitude of the expected turbulence.

We have also examined the possibility that the additional computation time of the ANN reconstructor will add to the latency of the AO system and hence diminish the performance. If the ANN induced significant performance degradation due to latency then we would expect to see the difference in performance between CARMEN and L\&A correlated with the atmospheric coherence time, $\tau_0$. The performance of CARMEN would be relatively worse for shorter $\tau_0$ than L\&A. Using turbulence strength and velocity measurements form Stereo-SCIDAR, Canary off-axis wavefront sensors and the local meteorological tower, and correcting for the airmass of the target asterism (greater airmass will increase the apparent wind speed if the wind direction is aligned with the target direction) we estimated $\tau_0$ for all of our observations. Values of $\tau_0$ ranged between approximately 3 and 9~ms. No correlation was found between the difference of the performance of CARMEN and L\&A and $\tau_0$. This is to be expected as, from section~\ref{sect:implem}, the computation time for CARMEN was estimated to be $1.01\pm0.01$~ms, significantly less than $\tau_0$.

Despite the limitations in the training of CARMEN, we achieve a performance within 5\% of the Strehl ratio of the optimised L\&A tomographic reconstructor. We have already shown in simulation \citep{Osborn12}, that with sufficient training data and negligible alignment errors the performance of CARMEN and L\&A are comparable. The next stage is to develop a training routine that can produce comparable results to L\&A whilst maintaining the generality that comes from the neural network approach.

\section{Conclusions}
\label{sect:conclusion}
We have shown that the Artificial Neural Network (ANN) open-loop tomographic reconstructor, CARMEN, is indeed insensitive to changes in the atmospheric optical turbulence profile. This was demonstrated on the Canary AO calibration bench. We see that the ANN provides a consistent reconstruction regardless of turbulence altitude without any additional information. There is a drop off in performance as the altitude of the layer increases which is consistent with the reduction of overlap of the projected pupils at that altitude.

We have also demonstrated that the CARMEN reconstructor, trained on the calibration bench, could attain results comparable to that of the Learn and Apply method. The performance was slightly lower than that of L\&A, with mean Strehl ratios of 0.31 and 0.26 for L\&A and ANN respectively. We believe that the lower performance was caused by insufficient data upon which the neural network was trained. This includes both the lack of variability in the phase screens that are used for generating the training dataset for the ANN reconstructor and the fact that there was turbulence above the altitude that the training data was acquired for. We maximised the phase variability in the training dataset by using two counter-rotating phase screens placed close to each other. This introduced a further issue that the two phase screens are actually displaced in altitude relative to each other. However, despite these limitations on the training of CARMEN we still achieve a performance within 5\% in terms of Strehl ratio and 15~nm in rms error of the optimised L\&A tomographic reconstructor. We also show that the reconstructor is performing a tomographic reconstructor as the performance is significantly better than that of GLAO in similar atmospheric conditions.

\section*{Acknowledgments}
DG appreciates support from CONICYT, through FONDECYT grant 11110149.
AG appreciates support from CONICYT, through FONDECYT grant 1120626.
FJdC appreciates support from the Spanish Economics and Competitiveness Ministry, through grant AYA2010-18513. This work was supported by Agence Nationale de la Recherche (ANR) program 06-BLAN-0191, CNRS / INSU, Observatoire de Paris, and UniversitŽ Paris Diderot Paris 7 in France, Science and Technology Facilities Council (ST/K003569/1 and ST/I002781/1), University of Durham in the UK and European Commission Framework Programme 7 (EELT Preparation Infrastructure Grant 211257 and OPTICON Research Infrastructures Grant 226604 and 312430). The Jacobus Kapteyn Telescope is operated on the island of La Palma by the Isaac Newton Group in the Spanish Observatorio del Roque de los Muchachos of the Instituto de Astrof'sica de Canarias. 
We would like to thank everyone who has been involved with Canary for designing and constructing an instrument which is ideal for testing many novel and exciting ideas.

\bibliographystyle{mn2e}

\end{document}